\documentclass[11pt,a4paper]{article}

\usepackage{amsmath,amssymb,epsfig,a4,latexsym,axodraw}
\usepackage{slashed}
\usepackage{jheppub}
\usepackage{subfigure}
\usepackage{etoolbox}

\allowdisplaybreaks
\newcommand{\RS}{\overline{\text{MS}}}
\newcommand{\DR}{\overline{\text{DR}}}
\newcommand{\Neps}{N_\epsilon}
\newcommand{\alphae}{\alpha_e}
\newcommand{\HiggsGlu}{\lambda}
\newcommand{\HiggsEps}{\lambda_{\epsilon}}
\newcommand{\alphaFourEps}{\alpha_{4\epsilon}}

\def\ghat{{\hat{g}}}
\def\gtilde{{\tilde{g}}}
\def\gbar{{\bar{g}}}
\def\HV{{\scshape hv}}
\def\FDH{{\scshape fdh}}
\def\DRED{{\scshape dred}}
\def\CDR{{\scshape cdr}}
\def\GammaAD{{\Gamma}}

\newcommand{\lnZ}[2]{\bar{#1}^{#2}}
\newcommand{\F}[3]{\bar{#1}^{#2}_{#3}}
\newcommand{\FFDH}[2]{\bar{F}^{#1}_{#2}}

\newcommand{\betaMSCDR}[1]{\beta^{\phantom{e}}_{#1}}

\newcommand{\betaMSsqCDR}[1]{\beta^{2}_{#1}}

\newcommand{\betaMS}[1]{\bar\beta^{\phantom{e}}_{#1}}
\newcommand{\betaeMS}[1]{\bar\beta^e_{#1}}
\newcommand{\betaMSsq}[1]{\bar\beta^{2}_{#1}}

\newcommand{\betaeDR}[1]{\bar\beta^{e,\overline{\text{DR}}}_{#1}}

\newcommand{\dZ}[2]{\delta\bar Z^{(#1)}_{#2}}
\newcommand{\gammaFDH}[2]{\bar\gamma^{#1}_{#2}}
\newcommand{\gammaDR}[2]{\bar\gamma^{#1,\overline{\text{DR}}}_{#2}}

\begin{document}
\thispagestyle{empty}
\begin{flushright}
PSI-PR-14-02\\
ZU-TH 08/14\\
\end{flushright}
\vspace{3em}
\begin{center}
{\Large\bf The Infrared Structure of QCD Amplitudes and\\[5pt] 
$H\to g g$ in FDH and DRED}
\\
\vspace{3em}
{\sc Christoph Gnendiger$^a$, Adrian Signer$^{b,c}$, Dominik St\"ockinger$^a$
}\\[2em]
{\sl ${}^a$ Institut f\"ur Kern- und Teilchenphysik,\\
TU Dresden, D-01062 Dresden, Germany\\
\vspace{0.3cm}
${}^b$ Paul Scherrer Institut,\\
CH-5232 Villigen PSI, Switzerland \\
\vspace{0.3cm}
${}^c$ Physik-Institut, Universit\"at Z\"urich, \\
Winterthurerstrasse 190,
CH-8057 Z\"urich, Switzerland}
\setcounter{footnote}{0}
\end{center}
\vspace{2ex}
\begin{abstract}
{} We consider variants of dimensional regularization, including the
four-dimensional helicity scheme (\FDH) and dimensional reduction
(\DRED), and present the gluon and quark form factors in the \FDH\,
scheme at next-to-next-to-leading order.
We also discuss the generalization of the infrared factorization formula
to \FDH\, and \DRED.
This allows us to extract the cusp anomalous dimension as well as the quark and gluon
anomalous dimensions at next-to-next-to-leading order in the \FDH\, and \DRED\, scheme,
using $\RS$ and $\DR$ renormalization.
To obtain these results we also present the renormalization procedure in these schemes.
\end{abstract}
\vspace{0.5cm}
\centerline

\newpage
\setcounter{page}{1}

\section{Introduction}
\label{sec:introduction}

The calculation of cross sections beyond leading order in perturbation
theory is of utmost importance to fully exploit the wealth of
experimental data provided by particle colliders. Computations at
next-to-leading order (NLO) are by now standard and can be done in
most cases in a fully automated way. At next-to-next-to-leading
order (NNLO) the situation is considerably more complicated and only a
small number of processes have been computed so far.

Beyond leading order, QCD cross sections are typically split into
several parts. At NLO there are virtual and real corrections, at NNLO
there are two-loop virtual, virtual-real and double real corrections.
Virtual corrections involve the calculation of loop diagrams and only
the sum of all contributions leads to finite results.

At intermediate steps of loop calculations ultraviolet (UV) and
infrared (IR) divergences need to be regularized.  Conventional
dimensional regularization (\CDR), where all vector bosons are treated
in $D=4-2\epsilon$ dimensions, is not always the optimal
choice. Alternatives are the 't Hooft-Veltman scheme (\HV)~\cite{'tHooft:1972fi},
the four-dimensional helicity (\FDH) scheme~\cite{BernZviKosower:1992}, and
dimensional reduction (\DRED) \cite{Siegel:1979wq}. In the latter two,
vector bosons are treated in 4 dimensions --- as far as possible.  As
an example of the use of the different schemes we mention the two-loop 
QCD results for the gluon-gluon and quark-gluon
scattering. Initially, the interference of these two-loop amplitudes
with the tree level was calculated in
\CDR~\cite{Glover:2001af, Anastasiou:2001sv}. Later the helicity
amplitudes were computed in the \HV\, and \FDH\,
scheme~\cite{Bern:2002tk, Bern:2003ck}. Clearly a full understanding
of the relation between the virtual corrections in the various schemes
is required if the \FDH\, or the \DRED\, scheme is to be used for the
computation of physical cross sections. Thus, the scheme dependence of
UV and IR singularities has to be studied.

The proper treatment of UV singlarities of pure QCD amplitudes in the
\FDH\, and \DRED\, scheme is well understood. The crucial step is to split
quasi-4-dimensional gluons into $D$-component gauge fields and
$\Neps=2\epsilon$ scalar fields, so-called $\epsilon$-scalars.
During the renormalization process the couplings of the $\epsilon$-scalars
must be treated as independent, resulting in different renormalization constants and
$\beta$-functions.
Ignoring this distinction can lead to wrong results, violation of unitarity,
and the non-cancellation of divergences~\cite{Kilgore:2011ta}
(see Ref.~\cite{Boughezal:2011br} for potential simplifications and 
alternative approaches).
The independent couplings and their renormalization were already
necessities in the equivalence proof of
\DRED\, and \CDR~\cite{Jack:1994bn, Jack:1993ws},
and in explicit multi-loop calculations in
\DRED~\cite{Harlander:2006rj, Harlander:2006xq, Harlander:2007ws}.

In non-supersymmetric theories the fact that we have different couplings
considerably complicates the renormalization procedure.
A case of particular interest is the gluon form factor,
i.e. the amplitude for the process Higgs to two gluons.
This process is described by an effective Higgs-gluon-gluon vertex including
the effective coupling $\lambda$ 
and has not been calculated at the two-loop level in \FDH\, or \DRED\, so far.
In these schemes there is an additional coupling $\lambda_\epsilon$
between the Higgs and two $\epsilon$-scalars and the renormalization
becomes highly non-trivial.

The split of gluons was also an essential ingredient in the resolution \cite{Signer:2005}
of the \DRED\, factorization problem \cite{Beenakker:1989,Smith:2005} and
lead to a better understanding of the one-loop transition rules of
Ref.\ \cite{Kunszt:1994}. It is clear that such a split has to be
the starting point for a consistent description of IR singularities in the
\FDH\, and \DRED\, scheme.  

In recent years a lot of progress has been made on the
understanding of the IR structure of gauge theories. In
Refs.~\cite{Gardi:2009zv, Becher:2009cu, Becher:2009qa}, a very simple
all-order formula predicting the IR divergences of pure QCD amplitudes
in \CDR\, has been proposed. An extension of this to the \FDH\, scheme,
based on Ref.~\cite{Gardi:2009zv}, 
has been presented by Kilgore~\cite{Kilgore:2012tb},
where transition rules for NNLO amplitudes computed in the \FDH\,
scheme to the \CDR\, (\HV) scheme were derived (for recent work on the
scheme dependence of double collinear splitting amplitudes
see~Ref.~\cite{Sborlini:2013jba}).  At one loop, the corresponding
transition rules~\cite{Kunszt:1994} can easily be realized by simple
scheme-dependent $\gamma^i$ constants for external partons $i$. Beyond
one loop the transition rules are more involved and
require a deeper understanding of IR singularities in loop amplitudes.

The aim of this paper is to deepen the understanding of the IR structure
of massless QCD amplitudes in \FDH\ and \DRED. We explain the
generalization of the IR prediction of Refs.\ \cite{Becher:2009cu,
  Becher:2009qa} for \CDR\ to the cases of \FDH\ and \DRED,
highlighting in particular the origin of the differences. As an
application and test we 
calculate the space-like two-loop form factors of quarks and gluons in
the \FDH\, and \DRED\, scheme.
We describe in detail the necessary UV renormalization
procedure in the $\RS$ and $\DR$ renormalization scheme and
extract the corresponding two-loop anomalous dimensions.

The structure of the paper is as follows:
After reminding the reader of the definitions of the various schemes
in Section~\ref{sec:dred}, we present a derivation of how to
extend the IR structure systematically to \FDH\, and \DRED\, in
Section~\ref{sec:infrared}.
The prediction of the IR structure is then tested in Section~\ref{sec:formfactors},
where we present the explicit two-loop results for the quark and
gluon form factors in \FDH. The renormalization procedure in general,
and the additional complications arising from the consistent renormalization of
the gluon form factor in the \FDH\, scheme in particular, are
discussed in Section~\ref{sec:renormalization}. With the help of these
results, in Section~\ref{sec:results} we are able to extract the cusp
anomalous dimension as well as the anomalous dimensions of quarks
and gluons at NNLO. These results are obtained in the
$\RS$ scheme, treating the $\epsilon$-scalars as independent particles
with multiplicity $N_\epsilon$. In Section~\ref{sec:DRbar} we then show
that the formalism also applies to the $\DR$ renormalization scheme, before
presenting our concluding remarks in Section~\ref{sec:conclusion}.

\section{QCD in different regularization schemes}
\label{sec:dred}

In all dimensional regularization schemes, momenta and space-time are
continued to $D=4-2\epsilon$ dimensions. UV and IR divergences of loop
and phase space integrals then appear as $1/\epsilon$ poles. In gauge
theories, such as QCD, the vector fields can be treated in different
ways. Following the detailed definitions in
Ref.\ \cite{Signer:2008va}, we distinguish four schemes: conventional
dimensional regularization (\CDR), the 't Hooft Veltman scheme (\HV),
the four-dimensional helicity scheme (\FDH), and regularization by
dimensional reduction (\DRED).

In \CDR\ and \HV, gluons are regularized in $D$ dimensions; the associated
$D$-dimensional metric tensor is denoted by $\ghat^{\mu\nu}$. In \FDH\,
and \DRED, gluons are regularized in $4$ dimensions; the associated
$4$-dimensional metric tensor is denoted by $g^{\mu\nu}$. Gauge
invariance on the regularized level requires that a $D$-dimensional
covariant derivative can be formed. Hence the $D$-dimensional space of
momenta must be a subspace of the 4-dimensional space of the
gluons. The metric tensors thus satisfy the relations
\begin{align}
g^{\mu\nu}&=\ghat^{\mu\nu}+\gtilde^{\mu\nu},&
g^{\mu\nu}\ghat_\nu{}^\rho&=\ghat^{\mu\rho},&
g^{\mu\nu}\gtilde_{\nu}{}^\rho&=\gtilde^{\mu\rho},&
\ghat^{\mu\nu}\gtilde_{\nu}{}^\rho&=0,\\
 g^{\mu\nu}g_{\mu\nu} &= 4,&
 \ghat^{\mu\nu}\ghat_{\mu\nu} &= D,&
\gtilde^{\mu\nu}\gtilde_{\mu\nu}&=2\epsilon,
\end{align} 
where a complementary $2\epsilon$-dimensional metric
$\gtilde^{\mu\nu}$ has been introduced.
Mathematical consistency requires \cite{Stockinger:2005gx} that this
``4-dimensional'' space cannot be the standard Minkowski space, but
must be realized as a more complicated space on which these metric
tensors can be defined. 

Not all gluons need to be regularized, but only internal ones, where
``internal'' gluons are defined as either virtual gluons that are part of a
one-particle irreducible loop diagram or, for real correction
diagrams, gluons in the initial or final state that are collinear or
soft. ``External gluons'' are defined as all other gluons. In
\CDR\, and \DRED, external gluons are treated in the same way as
internal ones. In \HV\, and \FDH, external gluons are not
regularized. In \FDH\, this implies that one needs to distinguish two
4-dimensional spaces---the one of the internal, regularized gluons
(metric $g^{\mu\nu}$) and the usual 4-dimensional Minkowski space
(metric $\gbar^{\mu\nu}$).
Table \ref{tab:RSs} summarizes the definitions of the four
regularization schemes.
 
In \FDH\, and \DRED, the (quasi-)4-dimensional regularized gluons can
be split into $D$-dimensional gluons (which appear in the
$D$-dimensional covariant derivative as gauge fields) and so-called
$\epsilon$-scalars with multiplicity $\Neps=2\epsilon$
\cite{Capper:1979ns}.  Often, this split is optional, but as discussed
in the introduction in some cases it is essential, see
Refs.~\cite{Jack:1994bn, Jack:1993ws, Harlander:2006rj,
  Harlander:2006xq, Harlander:2007ws, Signer:2005,Kilgore:2011ta}.  In
QCD with $N_F$ massless quarks we have to distinguish between
\begin{itemize}
\item the gauge coupling $\alpha_s$, appearing in all couplings of the
$D$-dimensional gluons,
\item the Yukawa-like evanescent coupling $\alphae$ between $\epsilon$-scalars
and quarks, and
\item the quartic $\epsilon$-scalar coupling $\alphaFourEps$. There are in
principle several independent such couplings, differing by the color
structure of the respective interactions, but in the present paper this
distinction is not necessary.
\end{itemize}

The renormalization is done by replacing the bare coupling constants
with the renormalized ones.
Most importantly, all couplings renormalize differently and the $\beta$-functions
for $\alpha_s$ and $\alphae$ needed in this paper are given by
\begin{subequations}
\begin{align}
 \mu^2\frac{\text{d}}{\text{d}\mu^2}\frac{\alpha_s}{4\pi}&=
 \betaMS{}(\alpha_s,\alphae,\epsilon)=
 -\epsilon\frac{\alpha_s}{4\pi}-
 \sum^{3}_{m+n}\betaMS{m n}
 \left(\frac{\alpha_s}{4\pi}\right)^m
 \left(\frac{\alphae }{4\pi}\right)^n
 +\mathcal{O}(\alpha^4)
 ,
 \label{eq:alphasRGE}\\
 \mu^2\frac{\text{d}}{\text{d}\mu^2}\frac{\alphae}{4\pi}&=
 \betaeMS{}(\alpha_s,\alphae,\epsilon)=
 -\epsilon\frac{\alphae}{4\pi}-
 \sum^{2}_{m+n}\betaeMS{m n}
 \left(\frac{\alpha_s}{4\pi}\right)^m
 \left(\frac{\alphae }{4\pi}\right)^n
 +\mathcal{O}(\alpha^3)
 .
 \label{eq:alphaeRGE}
\end{align}
\end{subequations}

The quartic coupling $\alphaFourEps$ does not appear at this level.
Here and in the following the bar denotes quantities obtained using \FDH\, or \DRED\, regularization.
In practical calculations, all couplings can be set numerically
equal, $\alpha_s=\alphae=\alphaFourEps$, but the $\beta$-functions and the 
related renormalization constants must be treated separately.
In contrast to this, in \CDR\, there is just the coupling $\alpha_s$ and we write
the well-known $\beta$-function as
\begin{align}
 \mu^2\frac{\text{d}}{\text{d}\mu^2}\frac{\alpha_s}{4\pi}&=
 \beta(\alpha_s,\epsilon)=
 -\epsilon\frac{\alpha_s}{4\pi}-
 \sum^{3}_{m}\beta_{m 0}
 \left(\frac{\alpha_s}{4\pi}\right)^m
 +\mathcal{O}(\alpha^4)
 .
 \label{eq:alphasRGECDR}
\end{align}

The starting point of the considerations in the next sections are known $\overline{\text{MS}}$
results of \CDR\, amplitudes. Because of this we use the following renormalization prescription
in the \FDH\, and \DRED\, scheme:
We treat $\epsilon$-scalars as independent scalar particles with an initially arbitrary
multiplicity $\Neps$. In the $\RS$ scheme we therefore subtract divergences of the form
$\left(\frac{\Neps}{\epsilon}\right)^{n}$.
As a consequence, $\betaMS{}$ and $\betaeMS{}$ depend
on the multiplicity $\Neps$ of the $\epsilon$-scalars,
and the value of the renormalized coupling $\alpha_s$
in this scheme equals the corresponding $\RS$ value in \CDR.

\begin{table}
\begin{center}
\begin{tabular}{l|cccc}
&\CDR&\HV&\FDH&\DRED\\
\hline
internal gluon&$\ghat^{\mu\nu}$&$\ghat^{\mu\nu}$&
$g^{\mu\nu}$&$g^{\mu\nu}$\\
 external gluon&$\ghat^{\mu\nu}$&$\gbar^{\mu\nu}$&
$\gbar^{\mu\nu}$&$g^{\mu\nu}$
\end{tabular}
\end{center}
\caption{
Treatment of internal and external gluons in the four different
regularization schemes,
i.e.\ prescription which metric tensor has to be used in propagator
numerators and polarization sums.
\label{tab:RSs}
}
\end{table}

\section{Infrared structure}
\label{sec:infrared}

On-shell scattering amplitudes in massless gauge theories remain
divergent even after UV renormalization. Fortunately, the remaining
infrared divergences factorize in a way that they can be absorbed by a
multiplicative renormalization, see
Refs.~\cite{Catani:1998bh,Sterman:2002qn,Gardi:2009zv,Becher:2009cu,Becher:2009qa}.

In the following we recapitulate the derivation of the factorization
formula in \CDR\, and show how it has to be modified in the cases of
\FDH\, and \DRED.

\subsection{CDR}
In the framework of dimensional regularization massless QCD amplitudes
with $n$ external partons 
can be written in the basis of an adequate color space as
\begin{align}
 \mathcal{M}_n\left(\epsilon,\frac{p_i}{\mu_r},\alpha_s(\mu_r)\right)=
 \mathbf{Z}\left(\epsilon,\frac{p_i}{\mu_f},\alpha_s(\mu_f)\right)
 H_n\left(\frac{p_i}{\mu_r},\frac{\mu_f}{\mu_r},\alpha_s(\mu_r)\right).
 \label{eq:factorization}
\end{align}

Here, $H_n$ denotes an arbitrary UV renormalized scattering amplitude,
which is finite in the limit $\epsilon\rightarrow 0$. Besides the
momenta of the external partons, $p_i$, and the running strong
coupling, $\alpha_s(\mu_r)$, it depends explicitly on the
renormalization scale, $\mu_r$, and the factorization scale, $\mu_f$.
To simplify things we set $\mu_r=\mu_f=\mu$ in the following.  All
soft and collinear divergences of $\mathcal{M}_n$ are combined in the
renormalization factor $\mathbf{Z}$.

In minimal subtraction schemes $\mathbf{Z}$ obeys a renormalization
group equation (RGE)
with a \textit{finite}, $\epsilon$-independent, anomalous dimension,
\begin{align}
 \frac{\text{d}}{\text{d\,ln}\,\mu}\mathbf{Z}\left(\epsilon,\frac{p_i}{\mu},\alpha_s(\mu)\right)
 =-\mathbf{\Gamma}\left(\frac{p_i}{\mu},\alpha_s(\mu)\right)\mathbf{Z}\left(\epsilon,\frac{p_i}{\mu},\alpha_s(\mu)\right),
 \label{eq:gammaRGE}
\end{align}
whose solution is given by the path ordered integral
\begin{align}
 \mathbf{Z}\left(\epsilon,\frac{p_i}{\mu},\alpha_s(\mu)\right)=
 -\mathcal{P}\,\text{exp}\int^{\mu}_0
 \frac{\text{d}\lambda}{\lambda}\,\mathbf{\Gamma}\left(\frac{p_i}{\lambda},\alpha_s(\lambda)\right).
 \label{eq:zSolution}
\end{align}

In Refs.~\cite{Gardi:2009zv,Becher:2009cu,Becher:2009qa} arguments are
put forward in favor of a conjecture for $\mathbf{\Gamma}$, which holds
at least up to the two-loop level:
\begin{align}
 \mathbf{\Gamma}\left(\frac{p_i}{\mu},\alpha_s(\mu)\right)=
 \sum^n_{(i,j)}\frac{\mathbf{T}_i\cdot\mathbf{T}_j}{2}\ \gamma^{\text{cusp}}\Big(\alpha_s(\mu)\Big)\ \text{ln} \frac{\mu^2}{-s_{ij}}+
 \sum_i \gamma^i\Big(\alpha_s(\mu)\Big).
 \label{eq:gammaConjecture}
\end{align}

The first sum describes the interaction of partons $i$ and $j$.
Due to large cancellations beyond
the one-loop level only two-particle interactions occur.  This term
contains the product $\mathbf{T}_i\cdot\mathbf{T}_j$ of the color
generators of partons $i$ and $j$, the kinematic variable
\mbox{$s_{ij}=\pm2\,p_i\cdot p_j$}, where the negative sign occurs if
not all momenta are incoming or outgoing, and the cusp anomalous
dimension $\gamma^{\text{cusp}}$.  The second sum represents the
collinear exchange of gluons and is given by the
anomalous dimensions $\gamma^i$ of all external partons $i$.  In
\CDR\, the anomalous dimensions $\gamma^{\text{cusp}}$ and $\gamma^i$
are known up to 3-loop order.

A direct consequence of the simple form of
\mbox{Eq.\ \eqref{eq:gammaConjecture}} is that the commutator
$[\mathbf{\Gamma}(\mu_1),\mathbf{\Gamma}(\mu_2)]$ vanishes and the
path ordering in \mbox{Eq.\ \eqref{eq:zSolution}} can be
neglected. Thus, the determination of $\mathbf{Z}$ reduces to a simple
integration of $\mathbf{\Gamma}$.  Here, one has to take into account
that the scale dependence of $\mathbf{\Gamma}$ is an explicit and
implicit one via the running of $\alpha_s$.  Because of this one first
has to solve the RGE \mbox{Eq.\ \eqref{eq:alphasRGECDR}} to express
$\alpha_s(\lambda)$ as a power series in $\alpha_s(\mu)$, and then
integrate \mbox{Eq.\ \eqref{eq:zSolution}}.  At this point it is
noteworthy that $\mathbf{\Gamma}$ itself does not depend explicitly on
the regularization parameter $\epsilon$.  The $\epsilon$-poles of
$\mathbf{Z}$ are a direct consequence of these two integrations.

Since the explicit scale dependence in
\mbox{Eq.\ \eqref{eq:gammaConjecture}} is a logarithmic one it is
useful to introduce the partial derivative of
$\mathbf{\Gamma}$
\begin{align}
 \Gamma'\Big(\alpha_s(\mu)\Big)
 =\frac{\partial}{\partial\,\text{ln}\,\mu}\mathbf{\Gamma}\left(\frac{p_i}{\mu},\alpha_s(\mu)\right)
 =-\,\gamma^{\text{cusp}}\Big(\alpha_s(\mu)\Big)\sum_i C_i.
 \label{eq:gammaPrime}
\end{align}
Here, the last equality follows from color conservation,
e.\,g. $\sum_i\mathbf{T}_i\mathcal{M}_n=0$, and
\mbox{$\mathbf{T}_i^2=C_i$}, where $C_i=C_{\bar q}=C_q=C_F$ for
(anti-)quarks and $C_i=C_g=C_A$ for gluons.

Now we specialize to the case of the space-like quark and gluon form factors,
where only two external colored partons appear.
Their momenta are normalized to $s_{12}=+2 p_1\cdot p_2=-1$
and the expansion in terms of the coupling $\alpha_s(\mu)$ reduces to
\begin{align}
 \mathbf{\Gamma}\left(\frac{p_i}{\mu},\alpha_s(\mu)\right)=
 \sum^{\infty}_{m=1}\left(\frac{\alpha_s}{4\pi}\right)^{m}\Big(\Gamma'_m\,\text{ln}\,\mu + \GammaAD_m\Big),
 \label{eq:gammaSimple}
\end{align}
with
\begin{subequations}
\begin{align}
 \label{eq:GammaPrimeCDR} \Gamma'_m &= -\,2\,\gamma^{\text{cusp}}_m\,
 C_{q/g},\phantom{\frac{1}{1}}\\ 
 \Gamma_m  &=+\,2\,\gamma^{q/g}_m.
 \label{eq:GammaCDR}
\end{align}
\end{subequations}

On the r.h.s. of Eq.~\eqref{eq:gammaSimple} and in the following the
argument of $\alpha_s(\mu)$ is suppressed.
Finally, Eq.~\eqref{eq:zSolution} yields for the case of form factors
\begin{align}
 \text{ln}\,\mathbf{Z}=
 \left(\frac{\alpha_s}{4\pi}\right)\left(\frac{\Gamma'_1}{4\epsilon^2}+\frac{\GammaAD_1}{2\epsilon}\right)+
 \left(\frac{\alpha_s}{4\pi}\right)^2\left(-\frac{3\beta_{20}\Gamma'_1}{16\epsilon^3}+
 \frac{\Gamma'_2-4\beta_{20}\GammaAD_1}{16\epsilon^2}+\frac{\GammaAD_2}{4\epsilon}\right)+
 \mathcal{O}(\alpha_s^3).
 \label{eq:lnZfinal}
\end{align}

Since
$\text{ln}\,\mathbf{Z}=\sum_m\left(\frac{\alpha_s}{4\pi}\right)^m\big(\text{ln}\,\mathbf{Z}\big)^{(m)}$
absorbs all infrared divergences of $\mathcal{M}_n$ the following
relations for the first coefficients hold:
\begin{subequations}
\begin{align}
 \Big(\text{ln}\,\mathbf{Z}\Big)^{(1)}&=\mathcal{M}^{(1)}_n\Big|_\text{poles},
 \label{eq:Zcoeff1}\\
 \Big(\text{ln}\,\mathbf{Z}\Big)^{(2)}&=\mathcal{M}^{(2)}_n\Big|_\text{poles}-\frac{1}{2}\left(\mathcal{M}^{(1)}_n\right)^2\Big|_\text{poles}.
 \label{eq:Zcoeff2}
\end{align}
\end{subequations}
With these formulas it is possible to determine the coefficients of
$\text{ln}\,\mathbf{Z}$ by a comparison with the IR pole structure of
UV renormalized amplitudes.

\subsection{FDH and DRED}

In the \FDH\, and \DRED\, scheme the logic of the derivation is
unchanged.  The crucial difference is that all quantities depend on
the additional couplings $\alphae$ and $\alphaFourEps$.  We stress
that although these two couplings are regularization artifacts the
behavior is the one of a gauge theory with scalar fields (whose
multiplicity happens to be $\Neps$) and with Yukawa-like and quartic
scalar interactions.

In the case of the renormalized two-loop quark and gluon form factors
the quartic coupling $\alphaFourEps$ does not appear and
the divergences can be absorbed by the modified renormalization factor,
\begin{align}
 \bar{\mathbf{Z}}\left(\epsilon,\frac{p_i}{\mu},\alpha_s(\mu),\alpha_e(\mu)\right)&=
 -\mathcal{P}\,\text{exp}\int^{\mu}_0
 \frac{\text{d}\,\lambda}{\lambda}\,\bar{\mathbf{\Gamma}}\left(\frac{p_i}{\lambda},\alpha_s(\lambda),\alpha_e(\lambda)\right).
 \label{eq:zSolutionDRED}
\end{align}

Likewise, the generalized anomalous dimension 
$\bar{\mathbf{\Gamma}}$ depends on the couplings $\alpha_s$ and $\alphae$:
\begin{align}
 \bar{\mathbf{\Gamma}}\left(\frac{p_i}{\mu},\alpha_s(\mu),\alpha_e(\mu)\right)&=
 \sum^n_{(i,j)}\frac{\mathbf{T}_i\cdot\mathbf{T}_j}{2}\ \gammaFDH{\text{cusp}}{}\Big(\alpha_s(\mu),\alphae(\mu)\Big)\ \text{ln} \frac{\mu^2}{-s_{ij}}+
 \sum_i \gammaFDH{i}{}\Big(\alpha_s(\mu),\alphae(\mu)\Big).
 \label{eq:gammaConjectureDRED}
\end{align}

Due to this, one has to solve Eqs.~\eqref{eq:alphasRGE} and
\eqref{eq:alphaeRGE} for $\alpha_s(\lambda)$ and $\alphae(\lambda)$,
respectively, before integrating Eq.~\eqref{eq:zSolutionDRED}.
Specializing again to the case of form factors and
expanding the result as a power series in $\alpha_s$ and $\alpha_e$ yields
\begin{align}
 \bar{\mathbf{\Gamma}}\left(\frac{p_i}{\mu},\alpha_s(\mu),\alpha_e(\mu)\right)
 &=\sum^{\infty}_{m+n=1}\left(\frac{\alpha_s}{4\pi}\right)^{m}\left(\frac{\alpha_e}{4\pi}\right)^{n}
 \Big(\bar{\Gamma}'_{m n}\,\text{ln}\,\mu + \bar\GammaAD_{m n}\Big),
 \label{eq:gammaSimpleDRED}
\end{align}
with
\begin{subequations}
\begin{align}
 \bar\Gamma'_{mn} &= - 2\, \gammaFDH{\text{cusp}}{mn}\, C_{q/g},
 \label{eq:GammaPrimeDRED}\phantom{\frac{1}{1}}\\
 \bar\Gamma_{mn}  &= + 2 \, \gammaFDH{q/g}{mn}.
 \label{eq:GammaDRED}
\end{align}
\end{subequations}

This leads to a modified expression for the renormalization factor,
\begin{align}
\begin{split}
 \text{ln}\,\bar{\mathbf{Z}}&=
 \left(\frac{\alpha_s}{4\pi}\right)\left(\frac{\bar\Gamma'_{10}}{4\epsilon^2}+\frac{\bar\GammaAD_{10}}{2\epsilon}\right)+
 \left(\frac{\alpha_e}{4\pi}\right)\left(\frac{\bar\Gamma'_{01}}{4\epsilon^2}+\frac{\bar\GammaAD_{01}}{2\epsilon}\right)\\
 &\quad+
 \left(\frac{\alpha_s}{4\pi}\right)^2
 \left(-\frac{3\betaMS{20}\bar\Gamma'_{10}}{16\epsilon^3}+
 \frac{\bar\Gamma'_{20}-4\betaMS{20}\bar\GammaAD_{10}}{16\epsilon^2}+\frac{\bar\GammaAD_{20}}{4\epsilon}\right)\\
 &\quad+
 \left(\frac{\alpha_s}{4\pi}\right)\left(\frac{\alpha_e}{4\pi}\right)
 \left(-\frac{3\betaeMS{11}\bar\Gamma'_{01}}{16\epsilon^3}+
 \frac{\bar\Gamma'_{11}-4\betaeMS{11}\bar\GammaAD_{01}}{16\epsilon^2}+\frac{\bar\GammaAD_{11}}{4\epsilon}\right)\\
  &\quad+
 \left(\frac{\alpha_e}{4\pi}\right)^2
 \left(-\frac{3\betaeMS{02}\bar\Gamma'_{01}}{16\epsilon^3}+
 \frac{\bar\Gamma'_{02}-4\betaeMS{02}\bar\GammaAD_{01}}{16\epsilon^2}+\frac{\bar\GammaAD_{02}}{4\epsilon}\right)+
 \mathcal{O}(\alpha^3).
 \label{eq:lnZfinalDRED}
\end{split}
\end{align}

Comparing this to Eq.~\eqref{eq:lnZfinal}, we notice that the
differences between the schemes are considerably more involved than at
the one-loop level.  Beyond one loop it is not possible any longer to
absorb all differences into shifts of the coefficients in
Eqs.~\eqref{eq:GammaPrimeCDR} and \eqref{eq:GammaCDR}. The additional
terms in Eq.~\eqref{eq:lnZfinalDRED} depend on the $\beta$-function
$\beta_e$ and/or the evanescent coupling $\alphae$ and have a much
more complicated structure.  However, as expected, in the limit
$\alphae \rightarrow 0$, Eq.~\eqref{eq:lnZfinalDRED} reduces to the
\CDR\, prediction.  The appearing $\beta$-functions can be taken from
the literature, see e.\,g. Refs.~\cite{Harlander:2006rj,
  Harlander:2006xq, Machacek:1983tz, Machacek:1983fi, Machacek:1984zw,
  Luo:2002ti}, and the only free parameters are the anomalous
dimensions $\bar\Gamma'_{ij}$ and $\bar\Gamma_{ij}$.  Again, they can
be determined by comparing the divergence structure with explicit
calculations, see Eqs.\ \eqref{eq:Zcoeff1} and \eqref{eq:Zcoeff2}.  In
the next section this is done for the space-like form factors of
quarks and gluons.

\section{Examples: Form factors of quarks and gluons in CDR and FDH}
\label{sec:formfactors}

The two-loop results of the quark and gluon form factors in \CDR\, are
known for quite some time~\cite{Gonsalves:1983nq, Harlander:2000mg}, and in fact
even the three-loop results are available~\cite{Gehrmann:2010ue}.
The divergent parts of the three-loop form factors in
\CDR~\cite{Moch:2005id,Moch:2005tm} have been used to extract the
anomalous dimensions $\gamma^q$~\cite{Becher:2006mr},
$\gamma^g$~\cite{Becher:2009qa}, and the cusp anomalous
dimension~\cite{Moch:2004pa} up to three-loop order.

In this section, we present the two-loop results of the quark and
gluon form factors obtained from an explicit calculation in the \FDH\,
scheme.  Since we are not considering contributions from external
$\epsilon$-scalars, this is equivalent to the \DRED\, scheme.  The
difference between the \CDR\, and \FDH\, results is due to diagrams
with internal $\epsilon$-scalars and, therefore, will also
involve the couplings $\alphae$ and $\alphaFourEps$.

To perform the calculations we used the following setup: the generation of the diagrams
and the implementation of the Feynman rules is done with the Mathematica package FeynArts~\cite{Hahn:2000kx};
the subsequent evaluation of the algebra in $D$ and $4$ dimensions is then performed with the package
TRACER~\cite{Jamin:1991dp}.
For the reduction and evaluation of the planar integrals we implemented an in-house algorithm
based on integration-by-parts methods and the Laporta-algorithm~\cite{Laporta:2001dd}.
The non-planar intagrals were reduced and evaluated with the packages
FIRE~\cite{Smirnov:2008iw} and FIESTA~\cite{Smirnov:2008py}, respectively.

\subsection{Quark form factor}
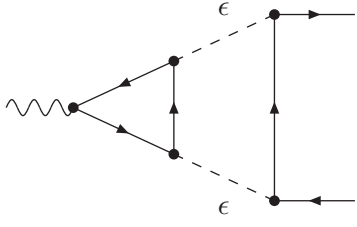
\begin{figure}[t]
\begin{center}
\begin{tabular}{ll}
\scalebox{1}{
\begin{picture}(135,90)(-5,-30)
\Vertex(25,45){2}
\Vertex(62.5,62.5){2}
\Vertex(62.5,27.5){2}
\Vertex(100,80){2}
\Vertex(100,10){2}
\Photon(0,45)(25,45){3}{3}
\ArrowLine(62.5,62.5)(25,45)
\ArrowLine(25,45)(62.5,27.5)
\ArrowLine(62.5,27.5)(62.5,62.5)
\DashLine(62.5,62.5)(100,80){4}
\DashLine(62.5,27.5)(100,10){4}
\ArrowLine(100,10)(100,80)
\ArrowLine(100,80)(130,80)
\ArrowLine(130,10)(100,10)
\Text(81.25,80)[b]{$\epsilon$}
\Text(81.25,5)[b]{$\epsilon$}
\end{picture}
}
\end{tabular}
\caption{\label{fig:tlcases}
Two-loop sample diagram resulting in a contribution $\propto\alphae^2$ to the quark form factor.
}
\end{center}
\end{figure}

At one loop, the quark form factor in \FDH\, receives additional
contributions $\propto\alpha_e$ from internal 
$\epsilon$-scalars coupling to quarks. Due to the Ward identity, no renormalization is required.
The explicit results in \CDR\, and \FDH, normalized to tree level, are denoted as
$F$ and $\bar F$, respectively. They read
\begin{subequations}
\begin{align}
\begin{split}
 F^{1l}_q(\alpha_s)&=
 \left(\frac{\alpha_s}{4\pi}\right) C_F \Bigg[-\frac{2}{\epsilon ^2}-\frac{3}{\epsilon }+\left(-8+\frac{\pi ^2}{6}\right)+
 \epsilon\left(-16+\frac{\pi ^2}{4}+\frac{14 \zeta(3)}{3}\right)\\
 &\quad\quad\quad\quad\quad\quad+\epsilon ^2\left(-32+\frac{2 \pi ^2}{3}+\frac{47 \pi ^4}{720}
 +7 \zeta (3)\right)\Bigg]+\mathcal{O}\left(\epsilon ^3\right),
 \end{split}
 \label{eq:Z1Quark}\\
 \begin{split}
 \F{F}{1l}{q}(\alpha_s,\alphae) & =
 F^{1l}_q(\alpha_s)+\left(\frac{\alphae}{4\pi}\right)
 \Neps C_F \Bigg[\frac{1}{2\epsilon}+\frac{1}{2}+\epsilon\left(\frac{1}{2}-\frac{\pi^2}{24}\right)
 \Bigg]
 +\mathcal{O}\left(\Neps\epsilon ^2\right).
 \end{split}
 \label{eq:Z1QuarkDRED}
\end{align}
\end{subequations}

The additional $\epsilon$-scalar contributions in Eq.~\eqref{eq:Z1QuarkDRED} are
proportional to $\alphae$ and $\Neps$.

Apart from contributions $\propto\alpha_s^2$, the two-loop quark form
factor in \FDH, $\F{F}{2l}{q}(\alpha_s,\alphae)$, contains also
terms $\propto\alpha_s \alphae$ and $\propto\alphae^2$. An example of
a diagram contributing to the latter is shown in
Figure~\ref{fig:tlcases}. Performing the explicit calculations in \CDR\,
and \FDH\, and forming the expressions relevant for
$\text{ln}\,\bar{\mathbf{Z}}$
we find
\begin{subequations}
\begin{align}
\begin{split}
 Q^{(2)}(\alpha_s)&\equiv
 F^{2l}_q(\alpha_s)-\frac{1}{2}\Big(F^{1l}_q(\alpha_s)\Big)^2\\
   &=
   \left(\frac{\alpha_s}{4\pi}\right)^2\Bigg\{
   C_A C_F \Bigg[\frac{11}{2 \epsilon ^3}+\frac{\frac{16}{9}+\frac{\pi ^2}{6}}{\epsilon ^2}-\frac{\frac{961}{108}+\frac{11\pi ^2}{12}-13 \zeta (3)}{\epsilon }\\
     &\quad\quad\quad\quad\quad\quad\quad\quad\quad-\frac{51157}{648}+\frac{11\pi^4}{45}-\frac{337\pi^2}{108}+\frac{313\zeta(3)}{9}\Bigg]\\
   &\quad\quad\quad\quad\quad+C_F^2\Bigg[\frac{-\frac{3}{4}+\pi ^2-12 \zeta (3)}{\epsilon }-\frac{1}{8}-\frac{11\pi^4}{45}+\frac{29\pi^2}{6}-30\zeta(3)\Bigg]\\
   &\quad\quad\quad\quad\quad+C_F N_F\Bigg[-\frac{1}{\epsilon ^3}-\frac{4}{9 \epsilon ^2}+\frac{\frac{65}{54}+\frac{\pi ^2}{6}}{\epsilon }+\frac{4085}{324}+\frac{23\pi^2}{54}+\frac{2\zeta(3)}{9}\Bigg]
   \Bigg\}\\
   &\quad+\mathcal{O}\left(\epsilon ^1\right),
 \end{split}
 \label{eq:Z2Quark}\\ \nonumber \\
 \begin{split}
 \lnZ{Q}{(2)}(\alpha_s,\alphae)&\equiv
 \F{F}{2l}{q}(\alpha_s,\alphae)-\frac{1}{2}\Big(\F{F}{1l}{q}(\alpha_s,\alphae)\Big)^2\\
 &=
 Q^{(2)}(\alpha_s)
   +\left(\frac{\alpha_s}{4\pi}\right)^2 N_{\epsilon } \Bigg\{C_A C_F
   \Bigg[-\frac{1}{4 \epsilon ^3}-\frac{1}{36 \epsilon ^2}+\frac{\frac{167}{216}+\frac{\pi ^2}{24}}{\epsilon }\Bigg]\Bigg\}\\
   &\quad+\left(\frac{\alpha_s}{4\pi}\right) \left(\frac{\alphae}{4\pi}\right) N_{\epsilon }\Bigg\{C_A C_F\frac{11}{4 \epsilon }+
   C_F^2\Bigg[-\frac{3}{2 \epsilon^2}-\frac{1+\frac{\pi ^2}{6}}{\epsilon }\Bigg]\Bigg\}\\
   &\quad+\left(\frac{\alpha_e}{4\pi}\right)^2 N_{\epsilon } \Bigg\{C_A C_F \Bigg[\frac{-\frac{1}{2}+\frac{N_{\epsilon }}{4}}{\epsilon ^2}\Bigg]
   +C_F^2 \Bigg[\frac{1}{\epsilon ^2}-N_{\epsilon }\Bigg(\frac{1}{4 \epsilon^2}+\frac{1}{16 \epsilon }\Bigg)\Bigg]\\
   &\quad\quad\quad\quad\quad\quad\quad+C_F N_F\Bigg[\frac{1}{4 \epsilon ^2}-\frac{3}{8 \epsilon }\Bigg]\Bigg\}+\mathcal{O}\left(\Neps\epsilon^0\right).
 \end{split}
 \label{eq:Z2QuarkDRED}
\end{align}
 \end{subequations}

Again, all additional terms in the \FDH\, result \eqref{eq:Z2QuarkDRED} are
proportional to at least one power of $\Neps$; in the contributions
proportional to $\alphae^2$ even $\Neps^2$ terms occur.  All results
have been obtained using $\overline{\rm MS}$ renormalization of
$\alpha_s$ and $\alpha_e$, Eqs.~\eqref{eq:alphasRGE} and \eqref{eq:alphaeRGE}.
The renormalization factors are listed in Section~\ref{sec:renormalization} for
convenience.

\subsection{Gluon form factor}

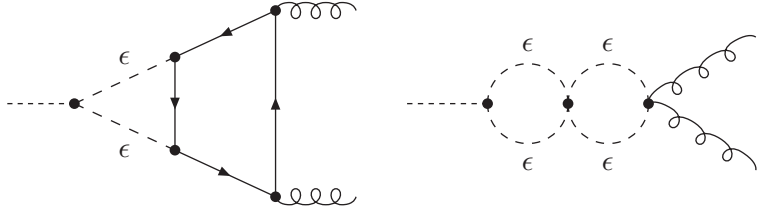
\begin{figure}[t]
\begin{center}
\begin{tabular}{ll}
\scalebox{1}{
\begin{picture}(135,90)(-5,-30)
\Vertex(25,45){2}
\Vertex(62.5,62.5){2}
\Vertex(62.5,27.5){2}
\Vertex(100,80){2}
\Vertex(100,10){2}
\DashLine(0,45)(25,45){2}
\DashLine(62.5,62.5)(25,45){4}
\DashLine(25,45)(62.5,27.5){4}
\ArrowLine(100,10)(100,80)
\ArrowLine(100,80)(62.5,62.5)
\ArrowLine(62.5,62.5)(62.5,27.5)
\ArrowLine(62.5,27.5)(100,10)
\Gluon(100,80)(130,80){3}{3}
\Gluon(130,10)(100,10){3}{3}
\Text(43.75,60)[b]{$\epsilon$}
\Text(43.75,25)[b]{$\epsilon$}
\end{picture}
\quad
\begin{picture}(135,90)(-5,-30)
\Vertex(30,45){2}
\Vertex(60,45){2}
\Vertex(90,45){2}
\DashCArc(45,45)(15,0,180){3}
\DashCArc(45,45)(15,180,360){3}
\DashCArc(75,45)(15,0,180){3}
\DashCArc(75,45)(15,180,360){3}
\DashLine(0,45)(30,45){2}
\Gluon(90,45)(130,70){3}{3}
\Gluon(90,45)(130,20){3}{3}

\Text(45,65)[b]{$\epsilon$}
\Text(45,20)[b]{$\epsilon$}
\Text(75,65)[b]{$\epsilon$}
\Text(75,20)[b]{$\epsilon$}
\end{picture}
}
\end{tabular}
\caption{\label{fig:tlcases2}
Sample diagram contributing to the gluon form factor $\propto\HiggsEps\alphae\alpha_s$
and $\HiggsEps\alphaFourEps\alpha_s$, respectively.
}
\end{center}
\end{figure}

The form factor of the gluon is computed in an effective theory
approach where the coupling $\HiggsGlu$ of the gluon to the Higgs is
induced through a dimension 5 operator. The renormalization of this
coupling in \CDR\, is well understood~\cite{Spiridonov:1984br}. In \FDH,
the presence of $\epsilon$-scalars induces an additional coupling to the Higgs,
$\HiggsEps$. This coupling is independent of $\HiggsGlu$ and
renormalizes differently. In fact the renormalization of $\HiggsGlu$ is
also affected by the presence of $\HiggsEps$.
In Section~\ref{sec:renormalization} we explain how to renormalize the
gluon form factor in the \FDH\, scheme.

After renormalization, the explicit results for $F^{1l}_g(\alpha_s) $
and $\F{F}{1l}{g}(\alpha_s,\HiggsEps/\HiggsGlu)$, the one-loop
gluon form factors in \CDR\, and \FDH, respectively, normalized to tree level, read
\begin{subequations}
\begin{align}
 \begin{split}
 F^{1l}_g(\alpha_s)&=
   \left(\frac{\alpha_s}{4\pi}\right)\Bigg\{ C_A\Bigg[
   -\frac{2}{\epsilon ^2}-\frac{11}{3 \epsilon }+\frac{\pi ^2}{6}+\epsilon\left(-2+\frac{14 \zeta (3)}{3}\right)\\
   &\quad\quad\quad\quad\quad\quad\quad+\epsilon ^2\left(-6+\frac{47 \pi ^4}{720}\right)\Bigg]
   +N_F\left(\frac{2}{3 \epsilon }\right)
   \Bigg\}
   +O\left(\epsilon ^3\right),
 \end{split}  
 \label{eq:Z1Gluon}\\
 \begin{split}
 \F{F}{1l}{g}(\alpha_s,\HiggsEps/\HiggsGlu) & =
   F^{1l}_g(\alpha_s)
   +\left(\frac{\alpha_s}{4\pi}\right)N_\epsilon C_A
   \Bigg\{\frac{1}{6\epsilon}
   +\frac{\HiggsEps}{\HiggsGlu}\left(1+3\epsilon
   \right)\Bigg\}
   +O\left(\Neps\epsilon ^2\right).
 \end{split}
 \label{eq:Z1GluonDRED}
\end{align}
\end{subequations}

All $\epsilon$-scalar terms in the \FDH\, result are proportional to $\alpha_s$ and $\Neps$.
The terms proportional to $\HiggsEps/\HiggsGlu$ appear from the ratio of the one-loop diagrams
$\propto\HiggsEps$, normalized to tree level.

At two loops, the gluon form factor in \FDH\, contains
also contributions $\propto\HiggsEps$, with some examples shown in
Figure~\ref{fig:tlcases2}. However, after renormalization and forming
the relevant expressions for $\text{ln}\,\bar{\mathbf{Z}}$
the contributions proportional to these couplings drop out, in
agreement with the IR prediction (\ref{eq:lnZfinalDRED}), which cannot contain the
coupling $\HiggsEps$.
The explicit results read
\begin{subequations}
\begin{align}
\begin{split}
G^{(2)}(\alpha_s)&\equiv
 F^{2l}_g(\alpha_s)-\frac{1}{2}\Big(F^{1l}_g(\alpha_s)\Big)^2\\
 &=
   \left(\frac{\alpha_s}{4\pi}\right)^2 \Bigg\{
   C_A^2\Bigg[\frac{11}{2 \epsilon^3}+\frac{3+\frac{\pi ^2}{6}}{\epsilon ^2}+\frac{-\frac{346}{27}+\frac{11 \pi ^2}{36}+\zeta (3)}{\epsilon }
   +\frac{5105}{162}+\frac{67\pi^2}{36}-\frac{143\zeta(3)}{9}\Bigg]\\
   &\quad\quad\quad\quad\quad
   +C_A N_F\Bigg[-\frac{1}{\epsilon ^3}-\frac{17}{9 \epsilon ^2}+\frac{\frac{64}{27}-\frac{\pi ^2}{18}}{\epsilon }
   -\frac{916}{81}-\frac{5\pi^2}{18}-\frac{46\zeta(3)}{9}\Bigg]\\
   &\quad\quad\quad\quad\quad
   +C_F N_F\Bigg[\frac{1}{\epsilon }-\frac{67}{6}+8\zeta(3)\Bigg]
   +N_F^2\frac{2}{9 \epsilon ^2}
   \Bigg\}
   +\mathcal{O}\left(\epsilon ^1\right),
\end{split}
 \label{eq:Z2Gluon}\\
\begin{split}
\lnZ{G}{(2)}(\alpha_s,\alphae)&\equiv
 \F{F}{2l}{g}(\alpha_s,\alphae,\HiggsEps/\HiggsGlu)-\frac{1}{2}\Big(\F{F}{1l}{g}(\alpha_s,\HiggsEps/\HiggsGlu)\Big)^2\\
 &=G^{(2)}(\alpha_s)
 +\left(\frac{\alpha _s}{4\pi}\right)^2 N_\epsilon\Bigg\{
   C_A^2\Bigg[-\frac{1}{4 \epsilon^3}+\frac{-\frac{7}{18}+\frac{N_{\epsilon }}{72}}{ \epsilon ^2}+\frac{\frac{49}{27}-\frac{\pi
   ^2}{72}}{\epsilon }\Bigg]
   +C_A N_F\frac{1}{9 \epsilon ^2}\Bigg\}
   \\&
   \quad
   +\left(\frac{\alpha_s}{4\pi}\right)\left(\frac{\alphae}{4\pi}\right) N_\epsilon\Bigg\{
   -\frac{C_F N_F}{2\epsilon}\Bigg\}
   +\mathcal{O}\left(\Neps\epsilon^0\right).
\end{split}
 \label{eq:Z2GluonDRED}
\end{align}
\end{subequations}

In contrast to the quark form factor, Eq.~\eqref{eq:Z2QuarkDRED}, the $\epsilon$-scalar terms in
Eq.~\eqref{eq:Z2GluonDRED} are much simpler and do not depend on $\alphae^2$.

\section{UV renormalization of the quark and gluon form factor in FDH}
\label{sec:renormalization}
   
Renormalization in the \FDH\, and \DRED\, scheme is considerably more
involved than in \CDR\ due to the additional evanescent
couplings. Here we present details on the renormalization in these
schemes, particularly for the gluon form factor, which involves not
only the renormalization of $\alpha_s$ and $\alpha_e$ but also of
composite operators and the associated couplings $\lambda$ and
$\lambda_\epsilon$.

In general, the renormalization of the quark and gluon form factors is
done by replacing the bare coupling constants with the renormalized
ones,
\begin{align}
 c_{\text{bare}}=c\left(1+\sum_{i}\delta Z^{(i)}_{c}\right), 
\end{align}
where $i$ indicates the loop order and $c\in\{\alpha_s,\HiggsGlu\}$ in
the case of \CDR\, and
$c\in\{\alpha_s,\alphae,\alphaFourEps,\HiggsGlu,\HiggsEps\}$ in \FDH.
As always, we use a bar to distinguish quantities in the \FDH\, scheme
from corresponding quantities in \CDR.

This leads to the following expressions for the coefficients of the
renormalized quark form factor in \CDR:
\begin{subequations}
\begin{align}
 F^{1l}_q(\alpha_s)&=
 F^{1l}_{q,\text{bare}}(\alpha_s),
 \label{eq:F1qRen}\\*
 F^{2l}_q(\alpha_s)&=
 F^{2l}_{q,\text{bare}}(\alpha_s)+ \delta Z^{(1)}_{\alpha_s}\,F^{1l}_{q,\text{bare}}(\alpha_s).
 \label{eq:F2qRen}
\end{align}
\end{subequations}

Due to the QED Ward-identity the photon coupling does not have to be
renormalized, and the bare and renormalized form factors are the same
at the one-loop level; at the two-loop level only the subloop
renormalization of $\alpha_s$ is necessary.

In \FDH, again no renormalization is needed at the one-loop level; at
the two-loop level the subloop renormalization of the couplings appearing in
the one-loop diagrams is necessary. Since all additional
$\epsilon$-scalar one-loop diagrams are proportional to $\alphae$, we
can write the \FDH\ renormalization as
\begin{subequations}
\begin{align}
 \FFDH{1l}{q}(\alpha_s,\alphae)&=
 \FFDH{1l}{q,\text{bare}}(\alpha_s,\alphae),
 \label{eq:F1qRenDRED}\\*
 \begin{split}
 \FFDH{2l}{q}(\alpha_s,\alphae)&=
 \FFDH{2l}{q,\text{bare}}(\alpha_s,\alphae)+ \dZ{1}{\alpha_s}\,F^{1l}_{q,\text{bare}}(\alpha_s)\\
 &\quad+\dZ{1}{\alphae}\,\left(\FFDH{1l}{q,\text{bare}}(\alpha_s,\alphae)-F^{1l}_{q,\text{bare}}(\alpha_s)\right).
 \label{eq:F2qRenDRED}
 \end{split}
\end{align}
\end{subequations}

Now we turn to the more complicated case of the gluon form factor.
Already at tree level it is proportional to the coupling $\HiggsGlu$,
which needs to be renormalized. Besides, the subloop renormalization
of both couplings appearing in the one-loop diagrams appears at higher
orders.  Thus, the \CDR\, coefficients of the renormalized gluon form
factor, normalized to tree level, read
\begin{subequations}
\begin{align}
 F^{1l}_g(\alpha_s)&=
 F^{1l}_{g,\text{bare}}(\alpha_s)+\delta Z^{(1)}_{\HiggsGlu},
 \label{eq:F1gRen}\\*
 F^{2l}_g(\alpha_s)&=
 F^{2l}_{g,\text{bare}}(\alpha_s)+\left(\delta Z^{(1)}_{\alpha_s}+\delta Z^{(1)}_{\HiggsGlu}\right)F^{1l}_{g,\text{bare}}(\alpha_s)
 +\delta Z^{(2)}_{\HiggsGlu}.
 \label{eq:F2gRen}
\end{align}
\end{subequations}

Renormalization in \FDH\ is more complicated because of the additional
coupling $\HiggsEps$ appearing in one-loop diagrams. Since the entire
one-loop difference between \FDH\ and \CDR\ is $\propto\HiggsEps\alpha_s$,
we can write
\begin{subequations}
\begin{align}
 \FFDH{1l}{g}(\alpha_s,\HiggsEps/\HiggsGlu)&=
 \FFDH{1l}{g,\text{bare}}(\alpha_s,\HiggsEps/\HiggsGlu)+\dZ{1}{\HiggsGlu},
 \label{eq:F1gRenDRED}\\*
 \begin{split}
 \FFDH{2l}{g}(\alpha_s,\alphae,\HiggsEps/\HiggsGlu)&=
 \FFDH{2l}{g,\text{bare}}(\alpha_s,\alphae,\alphaFourEps,\HiggsEps/\HiggsGlu)+
   \left(\dZ{1}{\alpha_s}+\dZ{1}{\HiggsGlu}\right)F^{1l}_{g,\text{bare}}(\alpha_s)\\
 &\quad+\left(\dZ{1}{\alpha_s}+\dZ{1}{\HiggsEps}\right)
 \left(\FFDH{1l}{g,\text{bare}}(\alpha_s,\HiggsEps/\HiggsGlu)-F^{1l}_{g,\text{bare}}(\alpha_s)\right)+\dZ{2}{\HiggsGlu}.
 \end{split}
 \label{eq:F2gRenDRED}
\end{align}
\end{subequations}
The couplings $\alphae$ and $\alphaFourEps$ only appear in two-loop
diagrams and don't have to be renormalized at this level.

The previous equations show which renormalization constants are needed
up to which order.  In \CDR, the required renormalization constants
read \cite{Spiridonov:1984br,Chetyrkin:1996ke,Chetyrkin:1997un}
\begin{subequations}
\begin{align}
 \delta Z^{(1)}_{\alpha_s}&=
 \Big(\frac{\alpha_s}{4\pi}\Big)\Big(-\frac{\betaMSCDR{20}}{\epsilon}\Big),\\*
 \delta Z^{(2)}_{\alpha_s}&=
 \Big(\frac{\alpha_s}{4\pi}\Big)^2\Big(\frac{\betaMSsqCDR{20}}{\epsilon^2}-\frac{\betaMSCDR{30}}{2\epsilon}\Big)
 ,\phantom{\frac{1}{1}}\\
 \delta Z^{(1)}_{\HiggsGlu}&=
 \delta Z^{(1)}_{\alpha_s}
 ,\phantom{\frac{1}{1}}\\
 \delta Z^{(2)}_{\HiggsGlu}&=
 \Big(\frac{\alpha_s}{4\pi}\Big)^2\Big(\frac{\betaMSsqCDR{20}}{\epsilon^2}-\frac{\betaMSCDR{30}}{\epsilon}\Big).
 \phantom{\frac{1}{1}}
\end{align}
\end{subequations}

Thus, the whole renormalization of the form factors is described by
the $\beta$-function of $\alpha_s$, defined in
Eq.~\eqref{eq:alphasRGECDR}, whose first non-vanishing coefficients in
the $\RS$ scheme are given by \cite{Harlander:2006rj,Harlander:2006xq}
\begin{subequations}
\begin{align}
 \betaMSCDR{20} &=\frac{11}{3}C_A-\frac{2}{3}N_F,
 \label{eq:beta20}\\
 \betaMSCDR{30} &=\frac{34}{3}C_A^2-\frac{10}{3}C_A N_F-2C_F N_F.
 \label{eq:beta30}
\end{align}
\end{subequations}

In the \FDH\ scheme, the additional $\epsilon$-scalar with
multiplicity $\Neps$ leads to a modification of the renormalization
constants for $\alpha_s$ and $\HiggsGlu$ and to new renormalization
constants for $\alphae$ and $\HiggsEps$.  The necessary
\FDH\ renormalization constants in the $\RS$ scheme described in
Section~\ref{sec:dred} read
\begin{subequations}
\begin{align}
 \dZ{1}{\alpha_s}&=
 \Big(\frac{\alpha_s}{4\pi}\Big)\Big(-\frac{\betaMS{20}}{\epsilon}\Big),
  \label{eq:deltaZ1alphas}\\
 \dZ{2}{\alpha_s}&=
 \Big(\frac{\alpha_s}{4\pi}\Big)^2\Big(\frac{\betaMSsq{20}}{\epsilon^2}-\frac{\betaMS{30}}{2\epsilon}\Big)
  +\Big(\frac{\alpha_s}{4\pi}\Big)\Big(\frac{\alphae}{4\pi}\Big)\Big(-\frac{\betaMS{21}}{2\epsilon}\Big)
 ,\phantom{\frac{1}{1}}
 \label{eq:deltaZ2alphas}\\
\dZ{1}{\HiggsGlu}&=
 \dZ{1}{\alpha_s}
 ,\phantom{\frac{1}{1}}
 \label{eq:deltaZ1HiggsGlu}\\
\dZ{2}{\HiggsGlu}&=
 \Big(\frac{\alpha_s}{4\pi}\Big)^2\Big(\frac{\betaMSsq{20}}{\epsilon^2}
 -\frac{\betaMS{30}}{\epsilon}\Big)
 +\Big(\frac{\alpha_s}{4\pi}\Big)\Big(\frac{\alphae}{4\pi}\Big)\Big(1+\frac{\HiggsEps}{\HiggsGlu}\Big)\Big(
 -\frac{\betaMS{21}}{2\epsilon}\Big)
 ,\label{eq:deltaZ2HiggsGlu}\\
 \dZ{1}{\alphae}&=
 \Big(\frac{\alpha_s}{4\pi}\Big)\Big(-\frac{\betaeMS{11}}{\epsilon}\Big)+
 \Big(\frac{\alphae }{4\pi}\Big)\Big(-\frac{\betaeMS{02}}{\epsilon}\Big),
 \label{eq:deltaZ1alphae}\\
 \dZ{1}{\HiggsEps}&=
 \Big(\frac{\alpha_s}{4\pi}\Big)\Big(-\frac{3C_A}{\epsilon}\Big)
 +\Big(\frac{\alphae}{4\pi}\Big)\frac{N_F}{\epsilon}
 +\Big(\frac{\alphaFourEps}{4\pi}\Big)C_A\Big(\frac{-1+N_\epsilon}{\epsilon}\Big),
\label{eq:deltaZ1HiggsEps}
\end{align}
\end{subequations}
with the following non-vanishing coefficients of the $\beta$-functions defined
in Eqs.~\eqref{eq:alphasRGE} and \eqref{eq:alphaeRGE}:
\begin{subequations}
\begin{align}
 \betaMS{20}&=\betaMSCDR{20} +\Neps\left(-\frac{C_A}{6}\right),
 \label{eq:beta20DRED}\\*
 \betaMS{30}&=\betaMSCDR{30} +\Neps\left(-\frac{7}{3}C_A^2\right),
 \label{eq:beta30DRED}\\*
 \betaMS{21}&=\Neps C_F N_F,\phantom{\frac{1}{1}}
 \label{eq:beta21DRED}\\
 \betaeMS{11}&=6\,C_F,\phantom{\frac{1}{1}}
 \label{eq:betae11}\\*
 \betaeMS{02}&=-4\,C_F+2\,C_A-N_F+\Neps\left(C_F-C_A\right).\phantom{\frac{1}{1}}
 \label{eq:betae20}
\end{align}
\end{subequations}

The modifications of the $\alpha_s$ and $\HiggsGlu$ renormalization
constants are of the order $\Neps$ and depends on all couplings including
$\alphae$ and $\HiggsEps$. The renormalization constant
$\dZ{1}{\HiggsEps}$ even depends on $\alphaFourEps$.

The renormalization of $\alpha_s$ and $\alphae$,
Eqs.~\eqref{eq:deltaZ1alphas}, \eqref{eq:deltaZ2alphas} and
\eqref{eq:deltaZ1alphae}, and all appearing $\beta$-functions are
obtained from
Refs.~\cite{Machacek:1983tz,Machacek:1983fi,Machacek:1984zw,Luo:2002ti},
where renormalization group equations for general gauge theories are
given.  We use the $\RS$ renormalization scheme described at the end
of Section~\ref{sec:dred}.  Extending the formalism described in
Ref.~\cite{Spiridonov:1984br} yields the renormalization of
$\HiggsGlu$, Eqs.~\eqref{eq:deltaZ1HiggsGlu} and
\eqref{eq:deltaZ2HiggsGlu}, including the appearance of $\HiggsEps$.
The renormalization of this coupling, Eq.~\eqref{eq:deltaZ1HiggsEps},
was obtained from an explicit one-loop calculation.

\section{Results: Anomalous dimensions in FDH and DRED}
\label{sec:results}

With the results from Section \ref{sec:formfactors} and
Eqs.~\eqref{eq:Zcoeff1}, \eqref{eq:Zcoeff2},
\eqref{eq:GammaPrimeDRED}, \eqref{eq:GammaDRED} and
\eqref{eq:lnZfinalDRED}
we are able to extract the scheme dependence of the
anomalous dimensions $\gamma^{\text{cusp}}$, $\gamma^{q}$ and
$\gamma^{g}$.  Here, the cusp anomalous $\gamma^{\text{cusp}}$ can be
extracted from both form factors, which allows for a cross check of
the method and the explicit calculation.

In the case of \CDR\ we recover the well-known results, see e.\,g. Ref.~\cite{Gehrmann:2010ue}
\begin{subequations}
\begin{align}
 \gamma_{10}^{\text{cusp}}&= 4,\phantom{\frac{1}{1}}
 \label{gc10}\\
 \gamma_{20}^{\text{cusp}}&= C_A\left(\frac{268}{9} - \frac{4}{3}\pi^2\right) - \frac{40}{9}N_F,\\
 \gamma_{10}^{q} & = -3\,C_{F},\phantom{\frac{1}{1}}\\
 \gamma_{20}^{q} & = C_{A} C_{F}\left(-\frac{961}{54}-\frac{11}{6}\pi^2+26\zeta(3)\right)
   +C_{F}^{2}\left(-\frac{3}{2}+2\pi^2-24\zeta(3)\right)\\ & \quad
   +C_{F} N_F\left(\frac{65}{27}+\frac{\pi^2}{3}\right),\\
 \gamma_{10}^{g}& = -\beta_{20} = -\frac{11}{3}C_A+\frac{2}{3}N_F,\\
 \gamma_{20}^{g}& = C_A^2\left(-\frac{692}{27}+\frac{11}{18}\pi^2+2\zeta(3)\right)+C_{A} N_F\left(\frac{128}{27}-\frac{\pi^2}{9}\right)+2C_{F} N_F
\end{align}
\end{subequations}

The additional contributions originating from internal
$\epsilon$-scalars in the \FDH\, or \DRED\, scheme lead to the following modified anomalous dimensions:
\begin{subequations}
\begin{align}
 \gammaFDH{\text{cusp}}{10}	&=\gamma_{10}^{\text{cusp}},\phantom{\frac{1}{1}}
 \label{gc10DRED}\\
 \gammaFDH{\text{cusp}}{01} 	&=0,\phantom{\frac{1}{1}}\\
 \gammaFDH{\text{cusp}}{20} 	&=\gamma_{20}^{\text{cusp}}-N_\epsilon\frac{16}{9} C_A,\\
 \gammaFDH{\text{cusp}}{11}	&=0,\phantom{\frac{1}{1}}\\
 \gammaFDH{\text{cusp}}{02}	&=0,\phantom{\frac{1}{1}}
\end{align}
\begin{align}
 & \gammaFDH{q}{10} = \gamma_{10}^{q},
 &&\gammaFDH{g}{10} = \gamma_{10}^{g} + N_\epsilon \frac{C_A}{6},\\*
 & \gammaFDH{q}{01} = N_\epsilon \frac{C_F}{2},
 &&\gammaFDH{g}{01} = 0,\\*
 & \gammaFDH{q}{20} = \gamma_{20}^{q} + N_\epsilon \Big(\frac{167}{108}+\frac{\pi^2}{12}\Big)C_A C_F,
 &&\gammaFDH{g}{20} = \gamma_{20}^{g} + N_\epsilon \Big(\frac{98}{27}-\frac{\pi^2}{36}\Big)C_A^2,\\*
 & \gammaFDH{q}{11} = N_\epsilon \Big[\frac{11}{2}C_A C_F-\Big(2+\frac{\pi^2}{3}\Big)C_F^2\Big],
 &&\gammaFDH{g}{11} =-\betaMS{21}=-\Neps C_F N_F,\\*
 &\gammaFDH{q}{02}  = - N_\epsilon \frac{3}{4} C_F N_F - N_\epsilon^2\frac{C_F^2}{8},
 &&\gammaFDH{g}{02} =0.
 \label{gg02DRED}
\end{align}
\end{subequations}

Generally, all these scheme differences are of 
$\mathcal{O}(\Neps)$ or $\mathcal{O}(\Neps^2)$,
so setting $\Neps$ to zero in
Eqs.~\eqref{gc10DRED}--\eqref{gg02DRED} yields the known \CDR\,
anomalous dimensions.
The one-loop cusp anomalous dimension obtained for both form factors
is scheme independent, while at two-loop order there is an additional
term in $ \gammaFDH{\text{cusp}}{20}$, i.e.\ 
a term proportional to $\alpha_s^2\Neps$.  The one-loop quark anomalous
dimension gets an additional contribution proportional to
$\alphae\Neps$, in the coefficient $\gammaFDH{q}{01}$, while the
$\alpha_s$ term is unchanged; at two-loop order, all coefficients
$\gammaFDH{q}{mn}$ get additional terms.
In the $\alphae^2$ part there is even a $\Neps^2$ term.
In the case of gluons the scheme dependence is absorbed by a term
$\propto\alpha_s\Neps$ at the one-loop level, and by terms
$\propto\alpha_s^2\Neps$ and $\propto\alpha_s\alphae\Neps$ at the
two-loop level 
(there are no terms $\propto \alphae^2$ and no terms containing $\alphaFourEps$).

Our results can be compared with Ref.~\cite{Kilgore:2012tb}, where the
gluon anomalous dimension has been obtained from the process $q\bar q\rightarrow g\gamma$.
As we are consistently using the $\RS$ scheme, as described at the end of Section~\ref{sec:dred},
the anomalous dimensions given here do not contain terms of $\mathcal{O}(\epsilon)$.
In  Ref.~\cite{Kilgore:2012tb}, such $\mathcal{O}(\epsilon)$ terms are
included to absorb process-specific contributions and lead to finite differences
for the two-loop anomalous dimensions.
Further, the $\mathcal{O}(\Neps^2)$ is missing in Ref.~\cite{Kilgore:2012tb}, which however
plays no role in the factorization formula~\eqref{eq:lnZfinalDRED} for $\Neps=2\epsilon$.

\section{Factorization in the $\overline{\text{DR}}$ scheme}
\label{sec:DRbar}

Up to now we considered a minimal coupling renormalization where all
additional UV singular contributions arising from internal
$\epsilon$-scalars are removed, including terms of the form
$\left(\frac{\Neps}{\epsilon}\right)^n$.  Now we show that the IR structure can be described
by Eq.~\eqref{eq:lnZfinalDRED} even if the $\DR$ renormalization scheme is used.

The $\DR$ scheme corresponds to setting $\Neps=2\epsilon$ and then subtracting
only the remaining $\frac{1}{\epsilon}$ UV poles. The difference between the $\RS$ and
$\DR$\, scheme are $\Neps$ terms in the $\beta$-functions and renormalization constants.

As it turns out, the structure of the factorization formula
\eqref{eq:lnZfinalDRED} is such, that an arbitrary $\Neps$ term in a
$\beta$ coefficient at the order $\mathcal{O}(\epsilon^{-n})$ can, for
$\Neps=2\epsilon$, be
absorbed by a \textit{finite} shift in the anomalous dimensions
$\Gamma'_{m n}$ and $\Gamma_{m n}$ at the order
$\mathcal{O}(\epsilon^{-n+1})$.  Since in Eq.~\eqref{eq:lnZfinalDRED}
no $\beta$-coefficients enter at the one-loop level, all corresponding
one-loop anomalous dimensions remain unchanged.  Comparing the
two-loop form factors renormalized in the $\overline{\text{DR}}$
scheme with the factorization formula we extract the two-loop
anomalous dimension in the $\overline{\text{DR}}$ scheme and find the
following results:
\begin{subequations}
\begin{align}
 \gammaDR{\text{cusp}}{10}&= \gamma_{10}^{\text{cusp}},\phantom{\frac{1}{1}}
 \label{gc10DREDDR}\\
 \gammaDR{\text{cusp}}{01}&= 0,\phantom{\frac{1}{1}}\\
 \gammaDR{\text{cusp}}{20}&= \gamma_{20}^{\text{cusp}}-\frac{4}{3}\,C_A
   ,\\
 \gammaDR{\text{cusp}}{11}&= 0,\phantom{\frac{1}{1}}\\
 \gammaDR{\text{cusp}}{02}&= 0,\phantom{\frac{1}{1}}
 \end{align}
 \begin{align}
 & \gammaDR{q}{10} = \gamma^{q}_{10},\phantom{\frac{1}{1}}
 &&\gammaDR{g}{10} = \gamma^{g}_{10},\phantom{\frac{1}{1}}\\
 & \gammaDR{q}{01} = 0,\phantom{\frac{1}{1}}
 &&\gammaDR{g}{01} = 0,\phantom{\frac{1}{1}}\\
 & \gammaDR{q}{20}  = \gamma_{20}^{q} + \frac{17}{9}\,C_A C_F,
 &&\gammaDR{g}{20} = \gamma_{20}^{g}+\frac{8}{9}\,C_A^2,\\
 %
 & \gammaDR{q}{11} = -\,\betaeDR{11}\,C_F,\phantom{\frac{1}{1}}
 &&\gammaDR{g}{11} = 0,\phantom{\frac{1}{1}}\\
 & \gammaDR{q}{02} = -\,\betaeDR{02}\,C_F ,\phantom{\frac{1}{1}}
 &&\gammaDR{g}{02} = 0,\phantom{\frac{1}{1}}
 \label{gg02DREDDR}
\end{align}
\end{subequations}
including the non-vanishing $\beta$-coefficients
\begin{subequations}
\begin{align}
 \betaeDR{11}&=\left.\betaeMS{11}\right|_{\Neps=0}
 =6\,C_F,\phantom{\frac{11}{3}}
 \label{eq:betaeDR11}\\*
 \betaeDR{02}&=\left.\betaeMS{02}\right|_{\Neps=0}
 =-4\,C_F+2\,C_A-N_F.
 \phantom{\frac{1}{1}}
 \label{eq:betaeDR02}
\end{align}
\end{subequations}

As expected, all one-loop quantities coincide with the corresponding $\overline{\text{MS}}$ values in \CDR\,
and the two-loop anomalous dimensions $\propto\alpha_s^2$ receive finite shifts.
Additionally, coefficients of the $\beta$-function $\betaeDR{}$ appear in the case of the two-loop quark form factor.
They are obtained from the previously used $\RS$ coefficients of $\betaeMS{}$ in the limit $\Neps=0$.

While in the case of the previous sections, the shifts in the anomalous dimensions were of the order 
$\mathcal{O}(\Neps)$, the $\gamma$-coefficients corresponding to $\DR$ renormalization differ by finite shifts,
which do not vanish for $\epsilon\rightarrow 0$.
This reflects the general fact that anomalous dimensions are renormalization-scheme dependent.

\section{Conclusion}
\label{sec:conclusion}

In this paper we extended the well-known \CDR\, conjecture \cite{Becher:2009cu,Becher:2009qa}
for the infrared structure of massless QCD amplitudes to the cases of \FDH\, and \DRED\,,
see Eq.~\eqref{eq:lnZfinalDRED}.
Consistently using the $\RS$ scheme, we extracted the NNLO anomalous dimensions by comparing
this conjecture with the form factors of quarks and gluons.
In the case of the gluon form factor we explained the necessary renormalization of the effective Higgs
couplings $\HiggsGlu$ and $\HiggsEps$.

In the $\RS$ scheme we treat the multiplicity $\Neps$ of the $\epsilon$-scalars as
an arbitrary quantity that enters in loop diagrams, and
the UV renormalization is done by subtracting all divergent parts, including terms of the
form $\left(\frac{\Neps}{\epsilon}\right)^n$.
The resulting regularization dependence can be absorbed in the modified infrared factorization
formula by unambiguously fixed shifts in the anomalous dimensions that are proportional
to at least one power of $\Neps$,  Eqs.~(\ref{gc10DRED})--(\ref{gg02DRED}).
Thus, after renormalization and after subtracting the corresponding IR divergent terms, 
the difference between an amplitude computed either in \CDR\, or \FDH\, is of
the order $\mathcal{O}(N_\epsilon)$ and free of $\frac{1}{\epsilon}$-poles.

This implies that the subtracted results in \CDR\, and \FDH\, are the
same for $N_\epsilon\to 0$ and it is possible to convert the results between the schemes.
The transition rules between \FDH\, and \CDR\, that follow from the anomalous dimensions given in
Eqs.~(\ref{gc10})--(\ref{gg02DRED}) are consistent with the transition rules given by
Kilgore~\cite{Kilgore:2012tb}.

Further we show how the \FDH\, and \DRED\, factorization works in other renormalization schemes,
namely in the $\DR$ scheme. Here, only remaining divergences after setting $\Neps=2\epsilon$
are subtracted. The resulting regularization dependence is absorbed by finite shifts
in the anomalous dimensions that do not depend on $\epsilon$ or $\Neps$,
Eqs.~(\ref{gc10DREDDR})--(\ref{gg02DREDDR}).
Thus, a transition to the \CDR\, anomalous dimensions like in the $\RS$ case is not possible.

In both renormalization schemes
the cusp anomalous dimension extracted from the quark form factor agrees with
the corresponding expression obtained from the gluon form factor.
This is further evidence for the universality of the proposed infrared 
structure in the \FDH\, and \DRED\, scheme.

In order to obtain transition rules for two-loop amplitudes in the
\DRED\, scheme, processes with external $\epsilon$-scalars need to be
considered. In particular, the corresponding anomalous dimension has
to be computed.  This can be done for example by computing the
$\epsilon$-scalar form factor corresponding to the process Higgs $\to$
two $\epsilon$-scalars. The investigation of alternative possibilities
to compute the anomalous dimensions more directly as well as the
application of the transition rules to the results for the $2\to 2$
scattering amplitudes in massless QCD~\cite{Glover:2001af,
  Anastasiou:2001sv,Bern:2002tk, Bern:2003ck} is left for future work.

\subsection*{Acknowledgements}
Communications with A. Broggio and A. Visconti are gratefully acknowledged.
This work has been supported by the German Research Foundation DFG through
Grant No. STO876/3-1.


\bibliography{bibliography}{}
\bibliographystyle{JHEP}
\end{document}